\documentclass[12pt]{iopart}

\usepackage{graphicx}
\usepackage{dcolumn}
\usepackage{color}
\usepackage{bm}

\usepackage{amsfonts}
\usepackage{amssymb}

\input epsf

\begin{document}

\title{Constraining alternative gravity theories using the solar neutrino problem}

\author{Sumanta Chakraborty}

\address{IUCAA, Post Bag 4, Ganeshkhind,\\
Pune University Campus, Pune 411 007, India}
\ead{sumantac.physics@gmail.com}

\pacs{$04.50.Kd;~14.60.Lm;~26.65.+t;~14.60.Pq$}

\begin{abstract}
The neutrino flavor oscillation is studied in some classes of
alternative gravity theories in a plane specified by $\theta =\pi
/2$, exploiting the spherical symmetry and general equations for
oscillation phases are given. We first calculate the phase in a
general static spherically symmetric model and then we discuss
some spherically symmetric solutions in alternative gravity
theories. Among them we discuss the effect of cosmological term in
Schwarzschild-(anti)de Sitter solution, which is the vacuum
solution in $F(R)$ theory, the effect of charge and Gauss-Bonnet
coupling parameter on the oscillation phase is presented. Finally
we discuss a charged solution with spherical symmetry in $F(R)$
theory and also its implication to the oscillation phase. We
calculate the oscillation length and transition probability in
these spherically symmetric spacetime and have presented a
graphical representation for transition probability with various
choice for parameters in our theory. From this we have constrained
parameters appearing in these alternative theories using standard
solar neutrino results.
\end{abstract}

\maketitle

\section{Introduction}\label{neuintro}

Neutrino oscillation is a very rich and interesting problem in its
own right. This problem is not only connected to modern particle
physics but also to cosmology, astrophysics and other diverse
branches, having interesting phenomenological consequences. Mass
neutrino mixing and oscillations were first studied and proposed
by Pontecorvo \cite{pont57}, then Mikheyev, Smirnov and
Wolfenstein (MSW for short) had discussed the effect of
transformation of one neutrino flavor into another in a varying
density medium \cite{mik86}, \cite{wolf78}. Recently, the mass
neutrino oscillation has been a hot topic and there have been many
theoretical (\cite{bil87}-\cite{sar88}) and as well as many
experimental (\cite{fuk98}-\cite{aha08}) studies. Neutrino
oscillations first formulated in flat spacetime has been extended
to curved spacetime (\cite{ahl96}-\cite{hua03}) and it has been
used to test equivalence principle recently \cite{man96}.
Oscillation phase along the geodesic line will produce a factor of
$2$ in the high energy limit when compared to the value along a
null geodesic. This factor of 2 exists in both flat and
Schwarzschild spacetime as shown by several authors (\cite{zha01},
\cite{bha99}-\cite{gro97}). The issue regarding this factor of $2$
is due to the difference between the time-like and null geodesics.
Also there exists some alternative mechanisms to take into account
the effect of gravitational field on Neutrino flavor oscillation
(\cite{gas88}, \cite{mur96}). Neutrino oscillation in non-inertial
frame has also drawn some attention recently (\cite{cap00},
\cite{lam01}). There have been extensive study of Neutrino
oscillation in spacetime with both curvature and torsion
(\cite{ali99}, \cite{cap99}).

In recent years, there have been a boost in the research on application of
Neutrino oscillation in various astrophysical contexts.
The Pulsar kicks mechanism, based on spin
flavor conversion of neutrino, which is propagating in a
gravitational field has been discussed extensively in \cite{lam05a}.
Observations suggest that pulsars have a very high
proper motion with reference to surrounding stars. These suggest that pulsars
undergo some kind of impulse or kick. In spite of many proposals in this direction this
still remain an open issue. The above pulsar kick mechanism can also be explained
by introducing \textit{neutrinospheres} and resonant oscillation $\nu _{e}\rightarrow \nu _{\mu ,\tau}$ between these
\textit{neutrinospheres}. This kind of Neutrino oscillation in presence of strong
magnetic field leads to such high proper motion of pulsars \cite{lam05b}.

Other studies on Neutrino oscillation mainly focusses on the fact
that though the mass squared differences and mixing angles are well observed, the absolute value
of neutrino masses are not properly known, leading to neutrino masses being hierarchical
or quasi-degenerate in nature. Along with there are extensive works on
the mixing angle $\theta _{13}$, CP violation in neutrino oscillation
and the effects of non vanishing $1-2$ mixing. There exists a number of theoretical models,
for example, considering neutrino masses to be degenerate at some seasaw scale, large mixing angle for
solar and atmospheric neutrinos using renormalization group equations in order to address these issues (\cite{lyc06}-\cite{akh08}).

There have been some recent developments regarding different
astrophysical aspects of alternative gravity theories
(\cite{cha12a}, \cite{cha12b}). In this paper we consider neutrino
oscillation in some classes of alternative gravity theories. For
simplicity we discuss only spherically symmetric solutions in
alternative theories of gravitation, but interestingly they all
turn out to have important implications. We have derived quiet
generally for all spherically symmetric solution of a particular
form (see equation (\ref{neu9})) that $\Phi _{k}^{geod}=2\Phi
_{k}^{null}$ and we have only taken the high energy limit but not
weak field approximation.

We have discussed three spherically symmetric solutions in this
paper. First one corresponds to vacuum solution to $F(R)$ gravity,
which is the Schwarzschild (anti-)de Sitter solution and have a
great importance today regarding the cosmological constant. We
have put an bound on the cosmological parameter in this solution
from the present day solar neutrino data. Secondly we consider
Einstein-Maxwell-Gauss-Bonnet (EMGB) gravity in five dimension and
a spherically symmetric solution has been discussed
\cite{Chakraborty}. There we put bounds on the GB parameter
$\alpha$ using solar neutrino oscillation data. Finally we discuss
charged solution in $F(R)$ gravity and different parameter have
been estimated.

Finally we calculate the proper oscillation length in all these
spherically symmetric spacetime. The oscillation length is found
proportional to $E_{loc}=E/\sqrt{g_{00}}$, the local energy
measurement. Decrease in local energy leads to decrease in
oscillation length as the neutrino travels out the gravitational
field. Thus blueshift of oscillation length occurs in contrast to
redshift for light signal, which is an interesting result.

The paper is organized as follows.
In section (\ref{neuflat}) we give a brief review of neutrino oscillation in flat spacetime,
next in section (\ref{neustat}) we discuss the neutrino oscillation in general static spherically symmetric spacetime.
Then we consider neutrino oscillation in different classes of alternative gravity theories and proper oscillation length in these theories.
The paper ends with a discussion on our results.
Throughout the paper we have used the units $G=c=\hbar =1$ and $\eta _{\mu \nu}=diag(+1,-1,-1,-1)$.

\section{Neutrino Oscillation in flat spacetime}\label{neuflat}

In this section we shall briefly review some properties of two
flavor neutrino oscillation in flat spacetime which will be
helpful for later developments. In standard treatment, the flavor
basis eigenstate, denoted by $\vert \nu _{\alpha}\rangle$ is
actually a superposition of the mass basis eigenstates $\vert \nu
_{k}\rangle$ such that they are connected by a unitary
transformation \cite{for97},
\begin{equation}\label{neu1}
\vert \nu _{\alpha}\rangle=\sum _{k}U_{\alpha k}~exp[-i\Phi _{k}]~\vert \nu _{k}\rangle
\end{equation}
where
\begin{equation}\label{neu2}
\Phi _{k}=E_{k}t-\overrightarrow{p_{k}}.\overrightarrow{x},~~~~~~~(k =1,2)
\end{equation}
and the unitary matrix $U_{\alpha k}$ comprises the transformation
between flavor and mass basis. Here $E_{k}$ and
$\overrightarrow{p_{k}}$ corresponds to the energy and momentum of
the mass eigenstates $\vert \nu _{k}\rangle$. For a neutrino which
is produced at some spacetime point,
$A(t_{A},\overrightarrow{x_{A}})$ and detected at another
spacetime point $B(t_{B},\overrightarrow{x_{B}})$, the phase as
presented in equation (\ref{neu2}) can be generalized to a
co-ordinate independent form and become suitable for application
in a curved spacetime. This could be given by
(\cite{for97},\cite{sto79}),
\begin{equation}\label{neu3}
\Phi _{k}=\int _{A}^{B}p_{\mu}^{(k)}dx^{\mu}
\end{equation}
where the 4-momentum is given by,
\begin{equation}\label{neu4}
p_{\mu}^{(k)}=m_{k}g_{\mu \nu}\frac{dx^{\nu}}{ds}
\end{equation}
and $m_{k}$ is the rest mass corresponding to the mass eigenstate
$\vert \nu _{k}\rangle$, $g_{\mu \nu}$ and $s$ corresponds to the
metric tensor and an affine parameter respectively. In the
literature the mass eigenstates are usually taken to be the energy
eigenstates with a common energy, up to ${\cal O} (m/E)$. We use
the approximation $E\gg M$ and assume massless trajectory implying
that the neutrino travels along the null trajectory. For two
flavor mixing $\nu _{e}-\nu _{\mu}$, we can write
\begin{equation}\label{neu5}
\nu _{e}=cos \theta \nu _{1}+sin \theta \nu _{2},~~~~~~~~~~~\nu _{\mu}=-sin \theta \nu _{1}+cos \theta \nu _{2}
\end{equation}
where $\theta$ is the vacuum mixing angle. The oscillation
probability that the neutrino which is produced as $\vert \nu
_{e}\rangle$ but detected as $\vert \nu _{\mu}\rangle$ is
given by \cite{boe92},
\begin{equation}\label{neu6}
P(\nu _{e}\rightarrow \nu _{\mu})=\vert \langle \nu _{e}\vert \nu _{\mu}(x,t)\rangle \vert ^{2}
=sin^{2}(2\theta)sin^{2}\left(\frac{\Phi _{kj}}{2}\right)
\end{equation}
where $\Phi _{kj}=\Phi _{k}-\Phi _{j}$ is the phase shift for
neutrino flavor oscillation. The Phase can also be expressed in
terms of energy and position of creation and detection of the
neutrino such that \cite{for97},
\begin{equation}\label{neu7}
\Phi _{k}\simeq m_{k}^{2}\vert \overrightarrow{x_{b}}-\overrightarrow{x_{A}} \vert (2E_{0})^{-1}
\end{equation}
with $E_{0}$ being the energy for a massless neutrino. So, the
phase shift which is responsible for oscillation is given by,
\begin{equation}\label{neu8}
\Phi _{kj}\simeq \Delta m_{kj}^{2}\vert \overrightarrow{x_{b}}-\overrightarrow{x_{A}} \vert (2E_{0})^{-1}
\end{equation}
where $\Delta m_{kj}^{2}=m_{k}^{2}-m_{j}^{2}$.
\section{Neutrino oscillation in a general static spherically symmetric spacetime}\label{neustat}

In this section we shall discuss the neutrino oscillation along
both null and timelike geodesics in a general static spherically
symmetric spacetime with metric ansatz \cite{cha11},
\begin{equation}\label{neu9}
ds^{2}=f(r)dt^{2}-f(r)^{-1}dr^{2}-r^{2}d\Omega _{2}^{2}
\end{equation}
We have restricted this discussion to four dimensions only, however it can be
generalized to higher dimension in a straightforward manner. We shall restrict our motion in
$\theta =\pi /2$ plane, due to spherical symmetry this would not hinder the general nature of the metric.

The components of the canonical momenta of $k$th massive neutrino in equation (\ref{neu4}) are,
\begin{eqnarray}\label{neu10}
p_{t}^{(k)}=p_{0}^{(k)}=m_{k}f(r)\dot{t}=m_{k}E_{k}
\nonumber
\\
p_{r}^{(k)}=-m_{k}f(r)^{-1}\dot{r}
\\
p_{\phi}^{(k)}=-m_{k}r^{2}\dot{\phi}=-m_{k}l_{k}
\nonumber
\end{eqnarray}
where we have introduced $\dot{t}=dt/ds$, $\dot{r}=dr/ds$ and $\dot{\phi}=d\phi /ds$.
The metric components do not depend on $t$, thus we have a conserved energy per particle mass given by
$E_{k}$ and the metric components also do not depend on $\phi$ leading to conserved angular momenta per
particle mass $l_{k}$. We could have $\dot{t}=E_{k}f(r)^{-1}$ and $\dot{\phi}=l_{k}r^{-2}$.
The phase along null geodesic from point $A$ to point $B$ is given by (\cite{for97},\cite{sto79}),
\begin{eqnarray}\label{neu11}
\Phi _{k}^{null}&=&\int _{A}^{B}p_{\mu}^{(k)}dx^{\mu}=\int _{A}^{B}\left(p_{0}^{(k)}dt + p_{\phi}^{(k)}d\phi + p_{r}^{(k)}dr\right)
\nonumber
\\
&=&\int _{A}^{B}\left(p_{0}^{(k)}dt/dr + p_{\phi}^{(k)}d\phi /dr + p_{r}^{(k)}\right)dr
\end{eqnarray}
In the literature the neutrino is usually taken to travel along the null line.
Thus we shall calculate the phase along light-ray trajectory from A to B.
The lagrangian appropriate for the motion in $\theta =\pi /2$ plane is,
\begin{equation}\label{neu12}
L=\frac{1}{2}\left(f(r)\dot{t}^{2}-f(r)^{-1}\dot{r}^{2}-r^{2}\dot{\phi}^{2}\right)
\end{equation}
The hamiltonian could be given by,
\begin{equation}\label{neu13}
H=E_{k}\dot{t}-l_{k}\dot{\phi} +m_{k}^{-1}p_{r}^{(k)}\dot{r}-L
\end{equation}
From independence of hamiltonian on time $t$ we can easily derive the following result,
\begin{equation}\label{neu14}
2H=E_{k}\dot{t}-l_{k}\dot{\phi} +m_{k}^{-1}p_{r}^{(k)}\dot{r}=\delta _{1}=constant
\end{equation}
We can take $\delta _{1}=1$ for time-like geodesics and $\delta _{1}=0$ for null geodesics without any loss of generality.
Substituting for $\dot{t}$, $\dot{\phi}$ and $p_{r}^{(k)}$ from equation (\ref{neu10}) in equation (\ref{neu14})
for null geodesics leads to the radial equation of motion,
\begin{equation}\label{neu15}
\dot{r}=E_{k}\sqrt{1-\frac{f(r)l_{k}^{2}}{r^{2}E_{k}^{2}}}
\end{equation}
Now we define a new function such that,
\begin{equation}\label{neu16}
V(r)=1-\frac{f(r)l_{k}^{2}}{r^{2}E_{k}^{2}}
\end{equation}
From this we have calculated the equations governing $t$ and
$\phi$ as,
\begin{equation}\label{neu17}
\frac{dt}{dr}=\frac{1}{f(r)\sqrt{V}},~~~~~~~~~~~~~~~~\frac{d\phi}{dr}=\frac{l_{k}}{E_{k}r^{2}\sqrt{V}}
\end{equation}
The on-mass shell condition corresponds to,
\begin{equation}\label{neu18}
m_{k}^{2}=g_{\mu \nu}p^{\mu}p^{\nu}=p_{0}^{(k)}p^{0}_{(k)}+p_{\phi}^{(k)}p^{\phi}_{(k)}+p_{r}^{(k)}p^{r}_{(k)}
\end{equation}
Using equation (\ref{neu10}) into the on-mass shell condition we readily obtain,
\begin{equation}\label{neu19}
p^{(k)r}=m_{k}\sqrt{E_{k}^{2}V-f(r)}
\end{equation}
Then using equations (\ref{neu19}), (\ref{neu17}), (\ref{neu15}) and (\ref{neu10}) in equation (\ref{neu11}) for phase we readily obtain,
\begin{equation}\label{neu20}
\Phi _{k}^{null}\simeq \int _{A}^{B}\frac{m_{k}dr}{2E_{k}\sqrt{V}}
\end{equation}
The phase as presented in equation (\ref{neu20}) is a general
result. For different $f(r)$ the function $V(r)$ changes and hence
the phase. If $f(r)=1-2M/r$, then in the high energy limit we
should have, $V\sim 1$, hence the phase has the following expression,
\begin{eqnarray}\label{neu21}
\Phi _{k}^{null}&=&\int _{A}^{B}\frac{m_{k}dr}{2E_{k}}
\nonumber
\\
&=&\frac{m_{k}^{2}}{2p_{0}^{k}}(r_{B}-r_{A})
\end{eqnarray}
which is the phase in Schwarzschild spacetime \cite{for97}. An
ultra relativistic neutrino travels with speed very close to that
of light and hence is considered to travel along the null line.
However there exists significant difference between massive
neutrino and photon, which becomes important while determining the
key features of neutrino oscillation. Thus for more general
situation we should calculate the phase along time-like geodesics.
An extra factor of 2 as mentioned earlier is obtained as we
compare the time-like geodesic with null geodesic in high energy
limit. This factor originates due to the fact that we have treated neutrino to be massive, while calculated the phase
along the null and the time-like trajectory. Thus this factor of 2 is
a consequence of neutrino mass. For time like geodesic, setting $\delta _{1}=1$, we can
derive from equation (\ref{neu14}),
\begin{equation}\label{neu22}
E_{k}\dot{t}-l_{k}\dot{\phi}+m_{k}^{-1}p_{r}^{(k)}\dot{r}=1
\end{equation}
Note that the equations for $\dot{t}$ and $\dot{\phi}$ are same for time-like geodesics \cite{ren10}. However the radial equation becomes,
\begin{equation}\label{neu23}
\dot{r}=\sqrt{E_{k}^{2}V-f(r)}
\end{equation}
Then we have obtained expressions for $dt/dr$ and $d\phi /dr$ for time-like geodesics given by,
\begin{equation}\label{neu24}
\frac{dt}{dr}=\frac{E_{k}}{f\sqrt{E_{k}^{2}V-f(r)}},~~~~~~~~~\frac{d\phi}{dr}=\frac{l_{k}}{r^{2}\sqrt{E_{k}^{2}V-f(r)}}
\end{equation}
Using the on-mass shell condition we readily obtain,
\begin{equation}\label{neu25}
p^{(k)r}=m_{k}\sqrt{E_{k}^{2}V-f(r)}
\end{equation}
Thus the phase along the time-like geodesic has the following expression
\begin{equation}\label{neu26}
\Phi _{k}^{geod}=\int _{A}^{B}\frac{m_{k}dr}{\sqrt{E_{k}^{2}V-f(r)}}
\end{equation}
In the high energy limit the above expression reduces to,
\begin{equation}\label{neu27}
\Phi _{k}^{geod}\simeq \int _{A}^{B}\frac{m_{k}dr}{E_{k}\sqrt{V}}=2\Phi _{k}^{null}
\end{equation}
The factor of $2$ exists in neutrino phase calculation for flat
\cite{lip00}, Schwarzschild (\cite{zha01},\cite{bha99}) and
Kerr-Newmann \cite{ren10} spacetime. Here we again found that
factor of $2$ for a general static spherically symmetric
spacetime. This factor appears since there exists intrinsic
difference between time-like and null geodesics. In deriving the
null phase we have used $4$-momentum which is along the time-like
geodesic and $\dot{r}$, along the null geodesic, however for
time-like phase we have derived both the quantities keeping them
along time-like geodesics, this leads to that factor of $2$, which
is a general feature of any curved spacetime.
\section{Neutrino Oscillation in Some Classes of Alternative Gravity Theories}\label{neualt}

Current theoretical models of cosmology have two fundamental
problems, namely inflation and the late time acceleration of the
universe. The usual scenarios used to explain both of these
accelerating epochs are to develop acceptable dark energy models,
which includes: scalar, spinor, cosmological constant and higher
dimensions. Even if such a model seems to be partially successful
it is mainly hindered by the coupling with the usual matter and
hence its compatibility with standard elementary particle
theories.

However another natural choice is the classical generalization of
general relativity, which is called modified gravity or
alternative gravity theory (\cite{cald03}, \cite{noj03},
\cite{noj07}, \cite{noj11}). Thus a gravitational alternative is
needed to explain both inflation and dark energy seems reasonable
on the ground of the expectation that general relativity is an
approximation valid at small curvature. The sector of modified
gravity theory which contains the gravitational terms, relevant at
high energy have produced the inflationary epoch. During evolution
the curvature decreases and hence general relativity describes to
a good approximation the intermediate universe. With a further
decrease of curvature as the sub-dominant terms gradually grow we
observe a transition from deceleration to cosmic acceleration.
There exists many models including traditional $F(R)$, string
inspired models, scalar tensor theories, Gauss-Bonnet theory and
many others. In the next subsections we shall discuss neutrino
oscillations in three spherically symmetric solutions for
different alternative gravity theories.

\subsection{Neutrino oscillation in $F(R)$ gravity}\label{neualt1}

General Relativity (GR) is widely accepted as one of the
fundamental theory relating matter energy density to geometric
properties of the spacetime. The standard cosmological model can
explain the evolution of the universe except inflation and late
time cosmic acceleration, as already mentioned. Although many
scalar field models have been proposed earlier in the frame work
of string theory and super-gravity to explain inflation however
Cosmic Microwave Background radiation does not show any evidence
in favor of some model. The same kind of approach has also been
taken to explain cosmic acceleration by introducing different dark
energy models where concrete observation is still missing.

Thus one of the simplest choice is modification of GR action by
introducing a term $F(R)$ in the lagrangian, where $F$ is some
arbitrary function of the scalar curvature $R$. There exists two
methods for deriving field equations, first, we can vary the
action with respect to metric tensor $g_{\mu \nu}$, the other
method which is called Palatini method is not discussed here. In
F(R) gravity (\cite{nel10}, \cite{cor10}, \cite{bal10},
\cite{fel10}), the Einstein-Hilbert action
\begin{equation}\label{va11}
S_{EH}=\int d^{4}x \sqrt{-g}\left(\frac{R}{16\pi} +L_{matter}\right),
\end{equation}
gets replaced by an action appropriate for the introduction of the
function of scalar curvature:
\begin{equation}\label{va12}
S_{F(R)}=\int d^{4}x \sqrt{-g}\left(\frac{F(R)}{16\pi} +L_{matter}\right)
\end{equation}
Varying this action we readily obtain the corresponding field
equation in this gravity theory to be given by,
\begin{equation}\label{va13}
\frac{1}{2}g_{\mu \nu}F(R)-R_{\mu \nu}F'(R)-g_{\mu \nu}\square F'(R)+\nabla _{\mu}\nabla _{\nu}F'(R) =-4\pi T_{matter \mu \nu}
\end{equation}
Now we shall discuss two class of solutions for the above set of
Einstein equations involving vacuum solution and charged black
hole solution in $F(R)$ gravity.
\subsubsection{Vacuum Solution in $F(R)$ gravity}\label{neuf1}

Several solutions (sometimes exact) to this field equation has been obtained, however due to complicated nature,
number of such exact solutions are much less than that in classical general relativity.
There exists a (A)dS-Schwarzschild solution that corresponds to a vacuum solution ($T=0$) for which the Ricci scalar is 
covariantly constant.
This also corresponds to $R_{\mu \nu}\propto g_{\mu \nu}$. Since $\square F'(R)=0$ for this case
equation ($\ref{va13}$) reduces to the following algebraic equation,
\begin{equation}\label{va14}
0=2F(R)-RF'(R)
\end{equation}
It is evident that the model $F(R) \propto R^{2}$ satisfy the above equation (\cite{noj11}, \cite{bal10}, \cite{fel10}).
Hence the (A)dS-Schwarzschild is an exact vacuum solution to this situation with respective line element given by,
\begin{equation}\label{va15}
ds^{2}=\left(1-\frac{2M}{r}\mp \frac{r^{2}}{L^{2}}\right)dt^{2}- \left(1-\frac{2M}{r}\mp \frac{r^{2}}{L^{2}}\right)^{-1}dr^{2}-r^{2}d\Omega ^{2}
\end{equation}
Here the minus(plus) sign corresponds to (anti-)de Sitter space,
$M$ is the mass of the black hole and $L$ is the length parameter of (anti-)de Sitter space,
which is related to the scalar curvature $R=\pm \frac{12}{L^{2}}$
(the plus sign corresponds to de Sitter space and minus sign corresponds to anti-de Sitter space).

The phase along null line is given by,
\begin{equation}\label{va16}
\Phi _{k}^{null}=\int _{A}^{B}\frac{m_{k}dr}{2E_{k}\sqrt{1-\frac{\left(1-\frac{2M}{r}\mp \frac{r^{2}}{L^{2}}\right)l_{k}^{2}}{r^{2}E_{k}}}}
\end{equation}
and the phase along the geodesic line has the following expression,
\begin{equation}\label{va17}
\Phi _{k}^{geod}=\int _{A}^{B}\frac{m_{k}dr}{\sqrt{E_{k}^{2}\left(1-\frac{\left(1-\frac{2M}{r}\mp \frac{r^{2}}{L^{2}}\right)l_{k}^{2}}{r^{2}E_{k}}\right)-\left(1-\frac{2M}{r}\mp \frac{r^{2}}{L^{2}}\right)}}
\end{equation}
\subsubsection{Charged Solution in $F(R)$ gravity}\label{neuf2}

In this section we consider charged solutions for $F(R)$ gravity having the form given by
$F(R)=R-\lambda exp(-\xi R)$ \cite{hen12}.
All such viable modifications in gravity must pass through all the tests from the
large scale structure of the universe to solar system.
When the correction factor to the Einstein gravity action is of exponential form,
it is possible to show that it does not contradicts solar system tests \cite{noji10}.
Also in addition the solutions from this model is mostly identical to that from Einstein
gravity except a change in Newton's constant \cite{Zha07}.
Using the function given by $F(R)$ we readily obtain topological
charged solution in which the function $f(r)$ as given by equation (\ref{neu9}) leads to \cite{hen12},
\begin{equation}\label{ch1}
f(r)=1-\frac{\Lambda}{3}r^{2}-\frac{M}{r}+\frac{Q^{2}}{r^{2}}
\end{equation}
In this solution we should set some parameters and following the approach
as presented in \cite{hen12} we easily read off the parameters such that the following relations are satisfied,
\begin{eqnarray}\label{ch2}
1+\frac{\lambda \xi}{e^{\xi R}}=0
\nonumber
\\
\frac{\lambda}{e^{\xi R}}+\frac{R}{2}\left(\frac{\lambda \xi}{e^{\xi R}}-1\right)=0
\end{eqnarray}
with the following solutions $\lambda =Re^{-1}$ and $\xi =-1/R$.
We can also reverse the argument i.e. setting $\lambda =Re^{-1}$ and $\xi =-1/R$ and deriving
that equation (\ref{ch1}) satisfies field equations.
To interpret the charge term we need scalar-tensor representation of $F(R)$ gravity theory.
Then the neutrino phase along the null line has the following expression,
\begin{equation}\label{ch3}
\Phi _{k}^{null}=\int _{A}^{B}\frac{m_{k}dr}{2E_{k}\sqrt{\left(1-\frac{f(r)l_{k}^{2}}{r^{2}E_{k}^{2}}\right)}}
\end{equation}
while that along the geodesic goes by the following expression,
\begin{equation}\label{ch4}
\Phi _{k}^{geod}=\int _{A}^{B}\frac{m_{k}dr}{\sqrt{E_{k}^{2}\left(1-\frac{f(r)l_{k}^{2}}{r^{2}E_{k}^{2}}\right)-f(r)}}
\end{equation}
Where $E_{k}$ and $l_{k}$ are the energy and angular momentum respectively of the neutrino and $f(r)$ is given by equation (\ref{ch1}).
\subsection{Neutrino Oscillation in Einstein-Maxwell-Gauss-Bonnet Gravity}\label{neualt2}

Theories with extra spatial dimension have been an area of
considerable interest since the original work of Kaluza and Klein.
The advent of string theory boosts this issue which predicts the
presence of extra spatial dimension. Among the large number of
alternatives the Brane world scenario is considered as a strong
candidate which has theoretical basis in some underlying string
theory. Usually, the effect of string theory on classical
gravitational theories (\cite{Green2}, \cite{Davies}) are
investigated using of a low energy effective action, which in
addition to the Einstein-Hilbert action contain squares and higher
powers of the curvature term. However the field equations become
of fourth order and brings in ghosts \cite{Zumino}. In this
context Lovelock \cite{Lovelock} showed that if higher curvature
terms appear in a particular combination in the action, the field
equation becomes of second order.

In Einstein-Maxwell-Gauss-Bonnet (EMGB) gravity, the action in the five dimensional spacetime ($M,g_{\mu \nu}$) can be written as,
\begin{equation}\label{46}
S=\frac{1}{2}\int _{M} d^{5}x \sqrt{-g} \left[R+\alpha L_{GB}+L_{matter} \right],
\end{equation}
where $L_{GB}=R_{\alpha \beta \gamma \delta}R^{\alpha \beta \gamma \delta}-4R_{\mu \nu}R^{\mu \nu}+R^{2}$ is
the GB Lagrangian and $L_{matter}=F^{\mu \nu}F_{\mu \nu}$ is the Lagrangian for the matter part i.e.
electromagnetic field. Here $\alpha$ is the coupling constant for the GB term having dimension of
$(length)^{2}$. As $\alpha$ is regarded as the inverse string tension, so we must have $\alpha \geq 0$.

The gravitational and electromagnetic field equations are obtained by
varying the above action with respect to $g_{\mu \nu}$ and $A_{\mu}$ (see \cite{Chakraborty}),
\begin{eqnarray}\label{47}
\left. \begin{array}{c}
G_{\mu \nu}-\alpha H_{\mu \nu}=T_{\mu \nu}\\
\bigtriangledown _{\mu}F^{\mu}_{\nu}=0\\
H_{\mu \nu}=2\left[RR_{\mu \nu}-2R_{\mu \lambda}R^{\lambda}_{\mu}-2R^{\gamma \delta}
R_{\mu \gamma \nu \delta}+R^{\alpha \beta \gamma}_{\mu}R_{\nu \alpha \beta \gamma} \right]-\frac{1}{2}g_{\mu \nu}L_{GB}
\end{array}\right \}
\end{eqnarray}
where $T_{\mu \nu}=2F^{\lambda}_{\mu}F_{\lambda \nu}-\frac{1}{2}F_{\lambda \sigma}F^{\lambda \sigma}g_{\mu \nu}$
is the electromagnetic field tensor.

A spherically symmetric solution to the above field equations has
been obtained by \cite{Dehghani} and the line element has the
following expression,
\begin{equation}\label{48}
ds^{2}=-g(r)dt^{2}+\frac{dr^{2}}{g(r)}+r^{2}d\Omega _{3}^{2},
\end{equation}
where the metric co-efficient is,
\begin{equation}\label{49}
g(r)=K+\frac{r^{2}}{4\alpha}\left[1\pm \sqrt{1+\frac{8\alpha \left(m+2\alpha \mid K \mid \right) }{r^{4}} -\frac{8\alpha q^{2}}{3r^{6}}} \right]
\end{equation}
Here $K$ is the curvature, $m+2\alpha \mid K\mid$ is the geometrical mass
of the spacetime and $d\Omega _{3}^{2}$ is the metric of a 3D hyper-surface such that,
\begin{equation}\label{50}
d\Omega _{3}^{2}=d\theta _{1}^{2}+sin^{2}\theta _{1}\left( d\theta _{2}^{2}+sin^{2}\theta _{2}d\theta _{3}^{2}\right)
\end{equation}
The range is given by $\theta _{1},\theta _{2}:[0,\pi]$.
We have assumed that there is a constant charge $q$ at $r=0$ and
the vector potential be $A_{\mu}=\Phi (r)\delta_{\mu}^{0}$ such that $\Phi (r)=-\frac{q}{2r^{2}}$.

In this metric the metric function $g(r)$ will be real for $r \geq r_{0}$ where $r_{0}^{2}$ is the largest real root of this cubic equation,
\begin{equation}\label{51}
3z^{3}+24\alpha \left(m+2\alpha \mid K \mid \right)z-8\alpha q^{2}=0
\end{equation}
By a transformation of the radial co-ordinates one can show that
$r=r_{0}$ is an essential singularity of the spacetime
\cite{Dehghani}.

The phase along the null line has the explicit form,
\begin{equation}\label{52}
\Phi _{k}^{null}=\int _{A}^{B}\frac{m_{k}dr}{2E_{k}\sqrt{\left(1-\frac{g(r)l_{k}^{2}}{r^{2}E_{k}^{2}}\right)}}
\end{equation}
and the phase along the geodesic line could be given by,
\begin{equation}\label{53}
\Phi _{k}^{geod}=\int _{A}^{B}\frac{m_{k}dr}{\sqrt{E_{k}^{2}\left(1-\frac{g(r)l_{k}^{2}}{r^{2}E_{k}^{2}}\right)-g(r)}}
\end{equation}
where $g(r)$ is given by equation (\ref{49}).
\section{Proper oscillation length}\label{neulength}

The propagation of a neutrino is well understood in terms of its proper length.
However the quantity $dr$ that appear in equation (\ref{neu20}) is only a coordinate.
The proper distance has the following expression \cite{lan87},
\begin{equation}\label{neul1}
dl=\sqrt{\frac{g_{0\mu}g_{0\nu}}{g_{00}}-g_{\mu \nu}dx^{\mu}dx^{\nu}}
\end{equation}
For the metric ansatz as presented in equation (\ref{neu9}) and then using equation (\ref{neu16})
we readily obtain the proper distance to be given by,
\begin{equation}\label{neul2}
dl=dr [f(r)V]^{-1/2}
\end{equation}
For convenience we shall adopt the differential form of (\ref{neu20})
\begin{equation}\label{neul3}
d\Phi _{k}^{null}=\frac{m_{k}dr}{2E_{k}\sqrt{V}}
\end{equation}
Substituting (\ref{neul2}) we readily obtain,
\begin{equation}\label{neul4}
d\Phi _{k}^{null}=\frac{m_{k}^{2}}{2p_{0}^{k}}\sqrt{f}dl
\end{equation}
Where it is generally assumed that the mass and energy eigenstates are identical with a common energy.
The equal energy assumptions are taken to be correct by some authors (\cite{zha01}, \cite{gro97})
and has been studied carefully in papers (\cite{leo00}). Also it is adopted quite widely in a
great deal of literature, that $p_{0}$ will represent the common energy of mass eigenstates.
The condition of equal momentum has also been adopted in order to study the neutrino oscillation.
In flat spacetime, both conditions represent the same neutrino oscillation results.
Due to the time and space translation invariance the free particle energy and momentum are conserved.
In curved, stationary spacetime, the energy is conserved along the geodesic due to existence
of a time-like Killing vector field. However $\partial /\partial r$ is not a Killing vector field
and hence the momentum $p_{r}$ is not conserved. Thus it is difficult to study neutrino oscillation
under the equal momentum assumption in a curved spacetime. In this section we shall consider phase along
the null line. Hence the phase shift determining the oscillation could be given by,
\begin{equation}\label{neul5}
d\Phi _{kj}^{null}=d\Phi _{k}^{null}-d\Phi _{j}^{null}=\frac{\Delta m_{kj}^{2}}{2p_{0}}\sqrt{f}dl
\end{equation}
where $\Delta m_{kj}^{2}=m_{k}^{2}-m_{j}^{2}$. Equation (\ref{neul5}) can be rewritten as,
\begin{equation}\label{neul6}
\frac{dl}{\left(d\frac{\Phi _{kj}^{null}}{2\pi}\right)}=\frac{4\pi p_{0}}{\Delta m_{kj}^{2}}\frac{1}{\sqrt{f(r)}}=
\frac{4\pi p_{o}^{loc}}{\Delta m_{kj}^{2}}
\end{equation}
The term $\frac{4\pi p_{0}}{\Delta m_{kj}^{2}}\frac{1}{\sqrt{g_{00}}}$ in (\ref{neul6})
has been interpreted as the oscillation length $L_{OSC}$ (which is actually defined by
the proper distance as the phase shift $\Phi _{kj}^{null}$ changes by $2\pi$) which is
measured by the observer at rest at a position $r$, and $p_{0}^{loc}=p_{0}/\sqrt{f(r)}$ is
being the local energy. As $r$ approaches to infinity, $p_{0}^{loc}$ approaches the energy
$p_{0}$ measured by an observer at infinity. Thus the neutrino oscillation length in a black
hole spacetime is given by following the metric ansatz (\ref{neu9}) as,
\begin{equation}\label{neul7}
L_{OSC}^{grav}=\frac{4\pi p_{0}}{\Delta m_{kj}^{2}}\frac{1}{\sqrt{f(r)}}
\end{equation}
while that for flat spacetime it reduces to,
\begin{equation}\label{neul8}
L_{OSC}^{flat}=\frac{4\pi p_{0}}{\Delta m_{kj}^{2}}
\end{equation}
Hence we can define an quantity which measures the fractional change in oscillation length due to presence of gravity,
\begin{equation}\label{neul9}
\delta l_{1}=\frac{L_{OSC}^{grav}-L_{OSC}^{flat}}{L_{OSC}^{flat}}=\frac{1}{\sqrt{f(r)}}-1
\end{equation}
Another quantity of interest is the shift in oscillation length due to these alternative theories compared
with the vacuum Schwarzschild solution in Einstein General Relativity and can be computed as,
\begin{equation}\label{neul10}
\delta l_{2}=\frac{L_{OSC}^{alter}-L_{OSC}^{sch}}{L_{OSC}^{flat}}=\frac{1}{\sqrt{f_{alt}(r)}}-\frac{1}{\sqrt{f_{sch}(r)}}
\end{equation}
where $f_{alt}(r)$ is the metric element for the alternative gravity theory
and $f_{sch}(r)$ is the metric element for schwarzschild theory i.e. $1-2M/r$. Next we shall calculate
these quantities for the spherically symmetric solution used previously in this paper and hence put bounds on the parameters.
\subsection{Vacuum Solution in $F(R)$ gravity}\label{neulv1}

\begin{figure}
\begin{center}
\begin{indented}

\includegraphics[height=2.6in, width=2.6in]{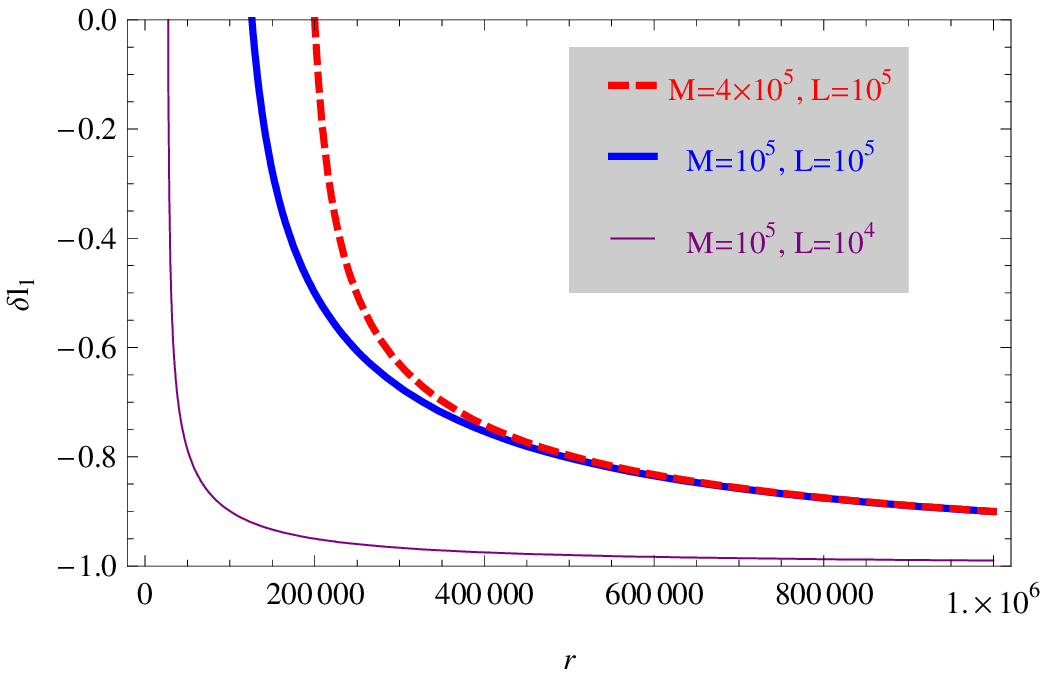}~~~~
\includegraphics[height=2.6in, width=2.6in]{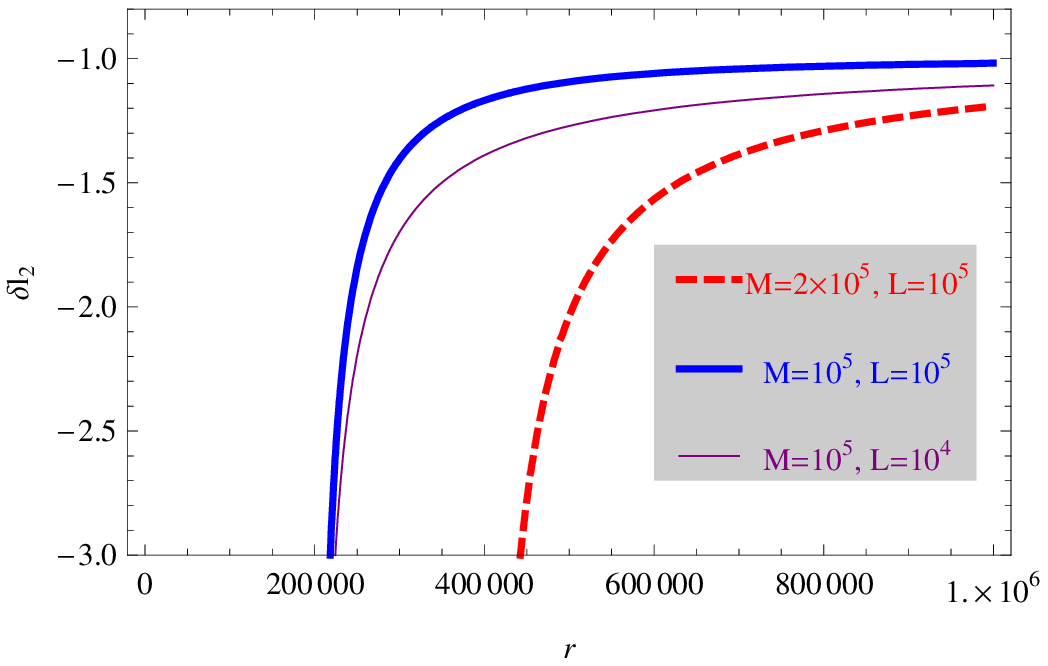}\\

\caption{Figure shows the variation of the two quantities $\delta
l_{1}$ and $\delta l_{2}$ defined in equations (\ref{neul11}) and (\ref{neul12}) for vacuum solution in $F(R)$ gravity with
radial coordinate for different choice the physical parameters i.e. $M$ and $L$. \label{fig1}
}

\end{indented}
\end{center}
\end{figure}

\begin{figure}
\begin{center}
\begin{indented}

\includegraphics[height=4in, width=4in]{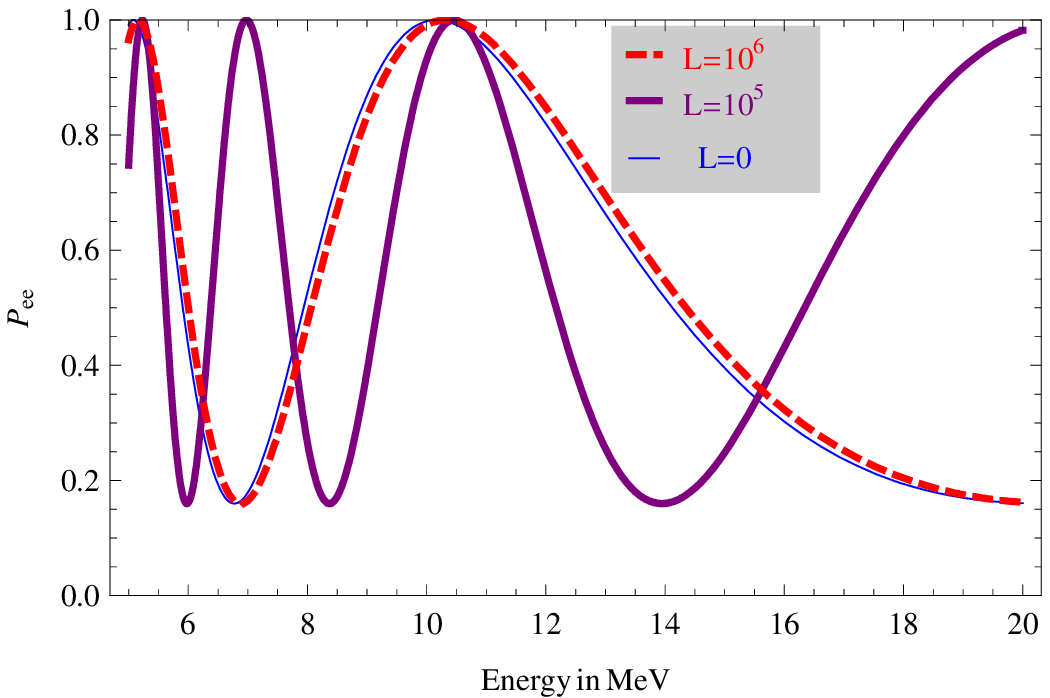}

\caption{The above figure illustrates the variation of $e\rightarrow e$
probability with neutrino energy measured in $MeV$ for an energy window. Different graphs are for different choice of $L$
in Vacuum solution for $F(R)$ gravity. The length is taken to be $180$ km.\label{fig2} }

\end{indented}
\end{center}
\end{figure}

We now consider proper oscillation length for neutrino oscillation in the vacuum solution for $F(R)$ gravity.
As pointed out in \ref{neuf1} the vacuum solution in $F(R)$ gravity actually comes form $F(R)\propto R^{2}$ and hence
completely different from the usual vacuum solution in Einstein theory for which $F(R)\propto R$.
Thus the vacuum solution in $F(R)$ gravity has the same structure
as (A)dS-Schwarzschild solution but is obtained from a $R^{2}$ lagrangian compared to $R$ in Einstein theory and differ from the standard
(A)dS-Schwarzschild solution in General Relativity (\cite{noj11}, \cite{bal10}, \cite{fel10}).
The quantities defined in equations (\ref{neul9}) and (\ref{neul10}) leads to the following expressions in this gravity theory,
\begin{equation}\label{neul11}
\delta l_{1}=\frac{1}{\sqrt{\left(1-\frac{2M}{r}\mp \frac{r^{2}}{L^{2}}\right)}}-1
\end{equation}
and,
\begin{equation}\label{neul12}
\delta l_{2}=\frac{1}{\sqrt{\left(1-\frac{2M}{r}\mp \frac{r^{2}}{L^{2}}\right)}}-\frac{1}{\sqrt{\left(1-\frac{2M}{r} \right)}}
\end{equation}
These two quantities are being plotted in figure \ref{fig1}, for the vacuum solution presented in this section.
From the figures we observe that $\delta l_{1}$ have the same asymptotic nature for all choice of parameters,
which is also valid for $\delta l_{2}$.
In this context we would like to constrain our parameters in the theory.
For this purpose we consider the oscillation probability of the neutrino to convert from one flavor to another.
For this purpose we use the data of solar neutrino oscillation, which is a two flavor
neutrino oscillation discussed in this paper. We present how the oscillation probability vary with the energy of the
neutrino for different choice of parameters. This variation of oscillation probability is presented in figure \ref{fig2}.

Now we present the data for solar neutrino in a tabular form and using the oscillation probability expression
we get bounds on the cosmological parameter $L$.
\begin{table}
\begin{center}
\begin{indented}

\caption{\bf$^{8}B$ solar neutrino results from real time experiments. The predictions of $BPS08(GS)$ and $SHP11(GS)$ standard
solar models are also shown. The errors are the statistical errors. Bounds on the cosmological parameter is estimated.}
\centering

\begin{tabular}{|c|c|c|c|}

\hline
\hline
{\bf Experiment} & {\bf Reaction} & {\bf $^{8}B$ $\nu$ flux} & {\bf Bound on} \\[0.3ex]

{} & {} & {} & {\bf cosmological parameter} \\[0.3ex]

{} & {} & {} & {$L^{-1}$}\\ [0.3ex]
\hline
\hline

Kamiokande \cite{kam}
&
$\nu e$
&
$2.80 \pm 0.19$
&
$< 2.235 \times 10^{-12}$\\

Super-K I \cite{supk}
&
$\nu e$
&
$2.38\pm 0.02$
&
$<2.325 \times 10^{-12}$\\

Super-K II \cite{supk2}
&
$\nu e$
&
$2.41 \pm 0.05$
&
$< 2.308 \times 10^{-12}$\\

Super-K III \cite{supk3}
&
$\nu e$
&
$2.32 \pm 0.04$
&
$<2.364 \times 10^{-12}$\\

SNO Phase I \cite{sno1}
&
CC
&
$1.76^{+0.06}_{-0.05}$
&
$<2.412 \times 10^{-12}$\\

~~~~(pure $D_{2}O$)
&
$\nu e$
&
$2.39^{+0.24}_{-0.23}$
&
$<2.323 \times 10^{-12}$\\

&
NC
&
$5.09^{+0.44}_{-0.43}$
&
$<2.029 \times 10^{-12}$\\

SNO Phase II \cite{sno2}
&
CC
&
$1.68 \pm 0.06$
&
$<2.423 \times 10^{-12}$\\

~~~~(NaCl in $D_{2}O$)
&
$\nu e$
&
$2.35 \pm 0.22$
&
$<2.328 \times 10^{-12}$\\

&
NC
&
$4.94 \pm 0.21$
&
$< 2.040 \times 10^{-12}$\\

SNO Phase III \cite{aha08}
&
CC
&
$1.67^{+0.05}_{-0.04}$
&
$<2.425 \times 10^{-12}$\\

~~~~~($^{3}He$ counters)
&
$\nu e$
&
$1.77 ^{+0.24}_{-0.21}$
&
$<2.411 \times 10^{-12}$\\

&
NC
&
$5.54^{+0.33}_{-0.31}$
&
$<1.895 \times 10^{-12}$\\

Borexino \cite{borx}
&
$\nu e$
&
$2.4 \pm 0.4$
&
$<2.312 \times 10^{-12}$\\

\hline

SSM [$BPS08(GS)$] \cite{bps08}
&
-
&
$5.94(1\pm 0.11)$
&
-\\

SSM [$SHP11(GS)$] \cite{shp11}
&
-
&
$5.58(1\pm 0.14)$
&
-\\

\hline
\hline

\end{tabular}
\end{indented}
\end{center}
\end{table}

\subsection{Proper Oscillation Length in Einstein-Maxwell-Gauss-Bonnet Gravity}\label{neuv2}

\begin{figure}
\begin{center}
\begin{indented}

\includegraphics[height=2.6in, width=2.6in]{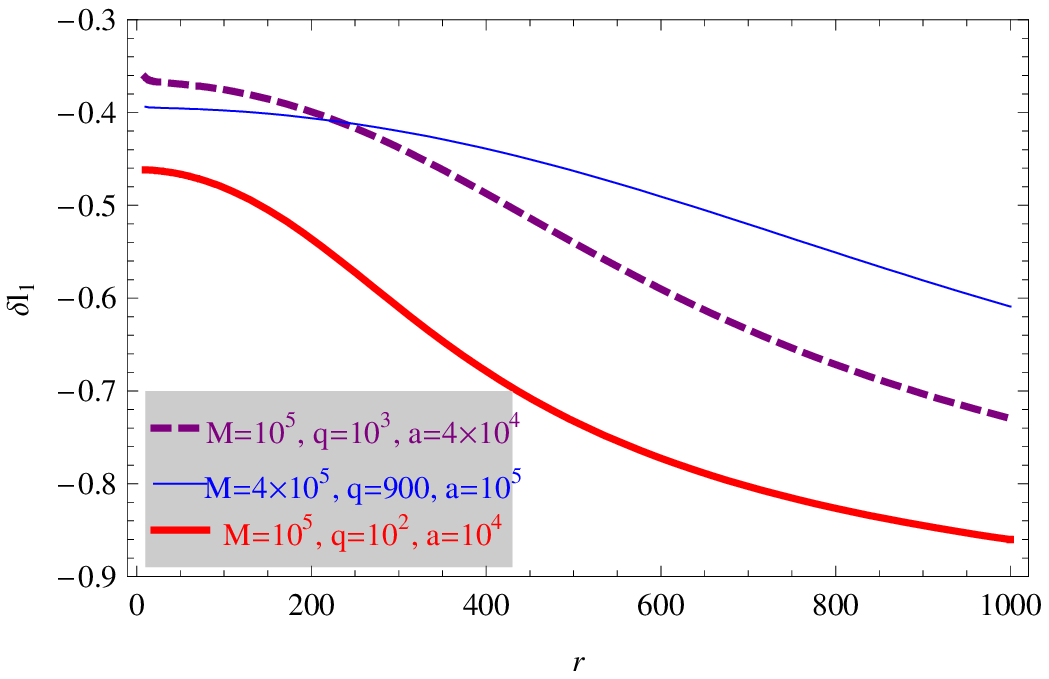}~~~~
\includegraphics[height=2.6in, width=2.6in]{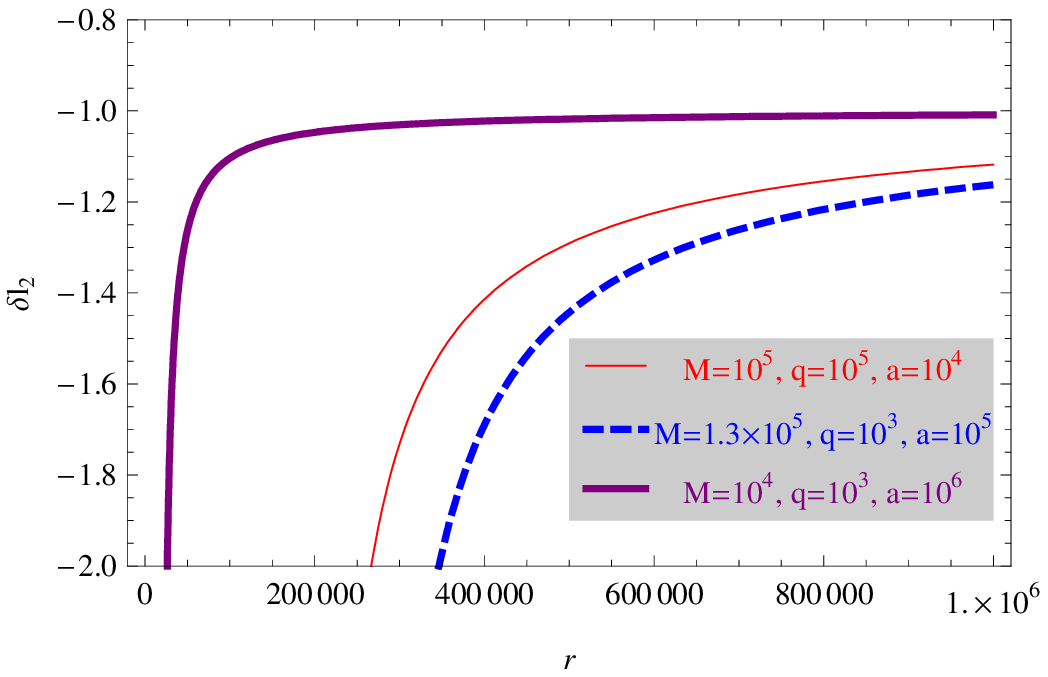}\\

\caption{Figures show the variation of the two quantities $\delta
l_{1}$ and $\delta l_{2}$ given by equations (\ref{neul13}) and (\ref{neul14})
for solution in EMGB gravity with
radial coordinate. Different graphs indicate different choices of parameters
in the theory, namely, $M$, $q$ and $\alpha$.\label{fig3} }

\end{indented}
\end{center}
\end{figure}

We consider proper oscillation length for neutrino oscillation in the EMGB gravity. The quantities defined in
equations (\ref{neul9}) and (\ref{neul10}) leads to the following expressions in this gravity theory given by,
\begin{equation}\label{neul13}
\delta l_{1}=\frac{1}{\sqrt{K+\frac{r^{2}}{4\alpha}
\left[1\pm \sqrt{1+\frac{8\alpha \left(m+2\alpha \mid K \mid \right) }{r^{4}} -\frac{8\alpha q^{2}}{3r^{6}}} \right]}}-1
\end{equation}
and,
\begin{equation}\label{neul14}
\delta l_{2}=\frac{1}{\sqrt{K+\frac{r^{2}}{4\alpha}\left[1\pm \sqrt{1+\frac{8\alpha \left(m+2\alpha \mid K \mid \right) }{r^{4}} -
\frac{8\alpha q^{2}}{3r^{6}}} \right]}}-\frac{1}{\sqrt{\left(1-\frac{2M}{r} \right)}}
\end{equation}
These two quantities defined above are being plotted in figure \ref{fig3}, for the charged solution presented in this section.
From the figures we observe that $\delta l_{1}$ and $\delta l_{2}$ have the same asymptotic behavior.

\begin{figure}
\begin{center}
\begin{indented}

\includegraphics[height=4in, width=4in]{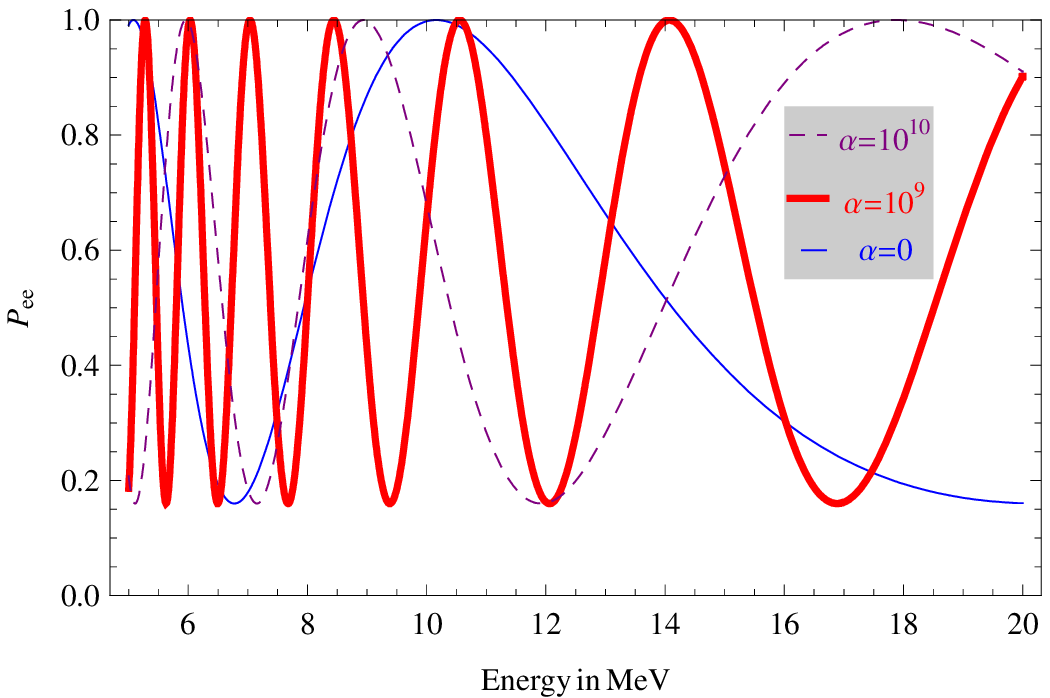}

\caption{Figure shows the variation of $e\rightarrow e$
probability with energy in $MeV$ for an energy window and length of $180$ km. Different graphs
are for different choices of parameter in EMGB gravity namely, $\alpha$. \label{fig4} }

\end{indented}
\end{center}
\end{figure}

We would like to constrain the string tension parameter $\alpha$ in the theory for some specific value of the charge
(the charge in the astrophysical situation being very small, so we have also taken charges in that order).
For this purpose we consider the oscillation probability of the neutrino to convert from one flavor to another as measured at earth.
The solution not being asymptotically flat has a non-negligible contribution even on the earth.
For this purpose we use the data of solar neutrino oscillation, which is a two flavor neutrino oscillation in gravitational
field discussed in this paper. This variation of oscillation probability is presented in figure \ref{fig4}.

Now the data for solar neutrino is presented in a tabular form and using the oscillation probability expression we get
bounds on the string tension $\alpha ^{-1}$.

\begin{table}
\begin{center}
\begin{indented}

\caption{\bf$^{8}B$ solar neutrino results from real time experiments. The predictions of $BPS08(GS)$ and $SHP11(GS)$ standard
solar models are also shown. The errors are the statistical errors. Bounds on string tension is estimated.}
\centering

\begin{tabular}{|c|c|c|c|}

\hline
\hline
{\bf Experiment} & {\bf Reaction} & {\bf $^{8}B$ $\nu$ flux} & {\bf Bound on} \\[0.3ex]

{} & {} & {} & {\bf string tension} \\[0.3ex]

{} & {} & {} & {$\alpha ^{-1}$}\\ [0.3ex]
\hline
\hline

Kamiokande \cite{kam}
&
$\nu e$
&
$2.80 \pm 0.19$
&
$< 5.935 \times 10^{-24}$\\

Super-K I \cite{supk}
&
$\nu e$
&
$2.38\pm 0.02$
&
$<6.213 \times 10^{-24}$\\

Super-K II \cite{supk2}
&
$\nu e$
&
$2.41 \pm 0.05$
&
$< 6.125 \times 10^{-24}$\\

Super-K III \cite{supk3}
&
$\nu e$
&
$2.32 \pm 0.04$
&
$<6.311 \times 10^{-24}$\\

SNO Phase I \cite{sno1}
&
CC
&
$1.76^{+0.06}_{-0.05}$
&
$<6.438 \times 10^{-24}$\\

~~~~(pure $D_{2}O$)
&
$\nu e$
&
$2.39^{+0.24}_{-0.23}$
&
$<6.313 \times 10^{-24}$\\

&
NC
&
$5.09^{+0.44}_{-0.43}$
&
$<5.892 \times 10^{-24}$\\

SNO Phase II \cite{sno2}
&
CC
&
$1.68 \pm 0.06$
&
$<6.523 \times 10^{-24}$\\

~~~~(NaCl in $D_{2}O$)
&
$\nu e$
&
$2.35 \pm 0.22$
&
$<6.281 \times 10^{-24}$\\

&
NC
&
$4.94 \pm 0.21$
&
$< 5.759 \times 10^{-24}$\\

SNO Phase III \cite{aha08}
&
CC
&
$1.67^{+0.05}_{-0.04}$
&
$<6.425 \times 10^{-24}$\\

~~~~~($^{3}He$ counters)
&
$\nu e$
&
$1.77 ^{+0.24}_{-0.21}$
&
$<6.312 \times 10^{-24}$\\

&
NC
&
$5.54^{+0.33}_{-0.31}$
&
$<5.891 \times 10^{-24}$\\

Borexino \cite{borx}
&
$\nu e$
&
$2.4 \pm 0.4$
&
$<6.391 \times 10^{-24}$\\

\hline

SSM [$BPS08(GS)$] \cite{bps08}
&
-
&
$5.94(1\pm 0.11)$
&
-\\

SSM [$SHP11(GS)$] \cite{shp11}
&
-
&
$5.58(1\pm 0.14)$
&
-\\

\hline
\hline

\end{tabular}
\end{indented}
\end{center}
\end{table}

\subsection{Proper Oscillation length in Charged Solution in $F(R)$ gravity}

\begin{figure}
\begin{center}
\begin{indented}

\includegraphics[height=2.6in, width=2.6in]{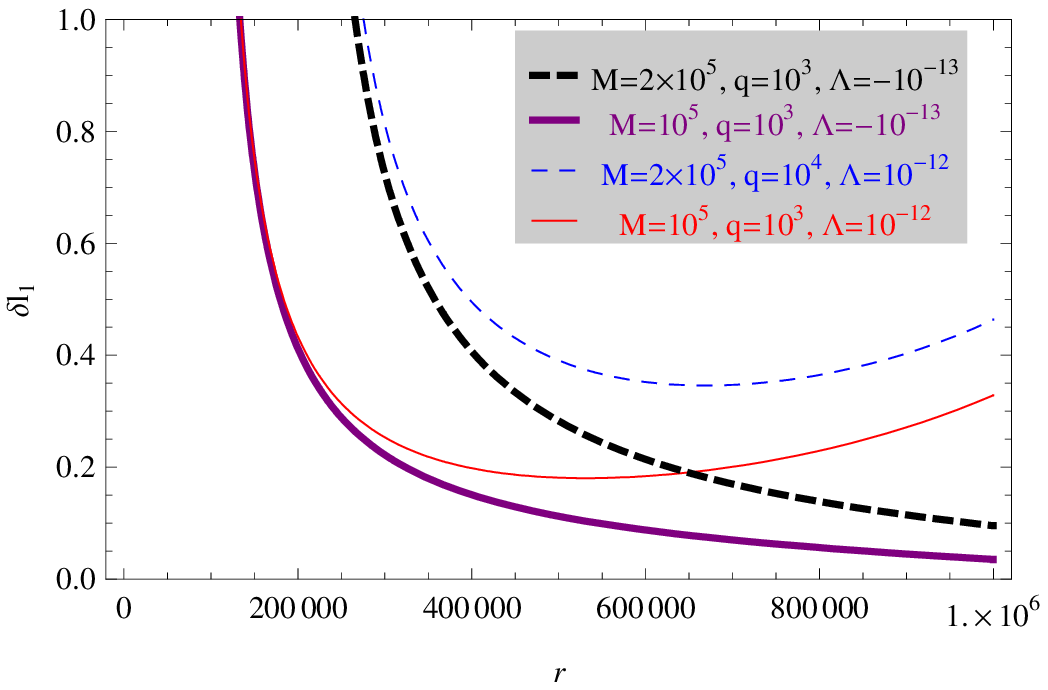}~~~~
\includegraphics[height=2.6in, width=2.6in]{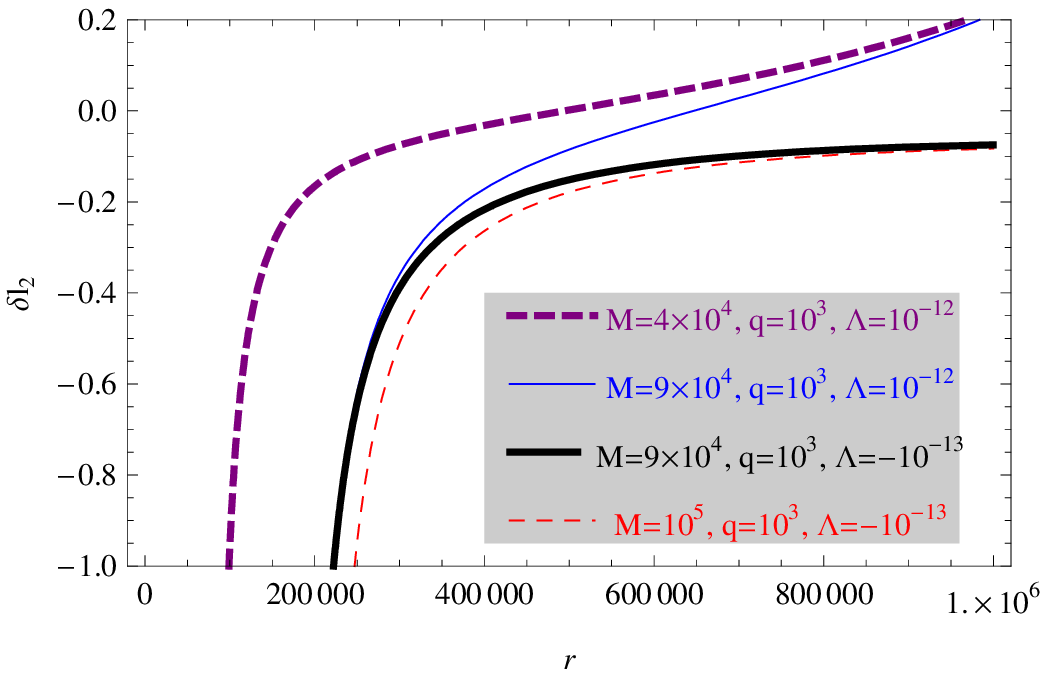}\\

\caption{Figure shows the variation of the two quantities $\delta
l_{1}$ and $\delta l_{2}$ as presented in equations (\ref{neul15}) and (\ref{neul16})
for charged solution in $F(R)$ gravity with
radial coordinate. Different graphs are for different choices of parameters namely,$M$, $q$ and $\Lambda$. \label{fig5} }

\end{indented}
\end{center}
\end{figure}

In this section proper oscillation length for neutrino oscillation in the charged $F(R)$ gravity has been discussed.
The quantities defined in equations (\ref{neul9}) and (\ref{neul10}) leads to the following expressions in this gravity theory such that,
\begin{equation}\label{neul15}
\delta l_{1}=\frac{1}{\sqrt{1-\frac{\Lambda}{3}r^{2}-\frac{M}{r}+\frac{Q^{2}}{r^{2}}}}-1
\end{equation}
and,
\begin{equation}\label{neul16}
\delta l_{2}=\frac{1}{\sqrt{1-\frac{\Lambda}{3}r^{2}-\frac{M}{r}+\frac{Q^{2}}{r^{2}}}}-\frac{1}{\sqrt{\left(1-\frac{2M}{r} \right)}}
\end{equation}
These two quantities defined above are being plotted in figure \ref{fig5}, for the charged solution presented in this section.
From the figures we observe that $\delta l_{1}$ and $\delta l_{2}$ have the same asymptotic behavior.

\begin{figure}
\begin{center}
\begin{indented}

\includegraphics[height=4in, width=4in]{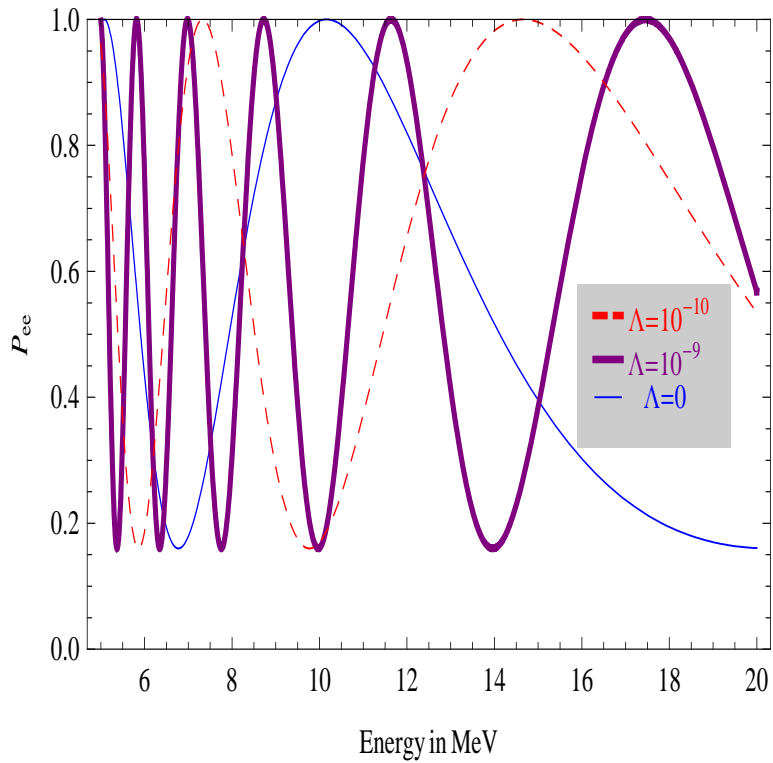}

\caption{The above figure shows the variation of $e\rightarrow e$
probability with energy in $MeV$ for a length of $180$ km. Different graphs are
for
different choices of parameters in charged $F(R)$ solution namely, $\Lambda$.\label{fig6} }

\end{indented}
\end{center}
\end{figure}

We would like to constrain the parameter $\Lambda$ in the theory for some specific value of the charge
(charge in the astrophysical situation being very small, so we have taken charges in that order).
For this purpose we consider the oscillation probability of the neutrino converting from one
flavor to another as measured on earth. For this purpose we use the data of solar neutrino oscillation,
which is a two flavor neutrino oscillation in gravitational field as discussed in this paper.
This variation of oscillation probability is presented in figure \ref{fig6}.

\begin{table}
\begin{center}
\begin{indented}
\caption{\bf$^{8}B$ solar neutrino results from real time experiments. The predictions of $BPS08(GS)$ and $SHP11(GS)$
standard solar models are also shown. The errors are the statistical errors. Bounds on the parameter $\Lambda$ is estimated.}
\centering

\begin{tabular}{|c|c|c|c|}

\hline
\hline
{\bf Experiment} & {\bf Reaction} & {\bf $^{8}B$ $\nu$ flux} & {\bf Bound on parameter} \\[0.3ex]

{} & {} & {} & {\bf in charged theory} \\[0.3ex]

{} & {} & {} & {$| \Lambda |$}\\ [0.3ex]
\hline
\hline

Kamiokande \cite{kam}
&
$\nu e$
&
$2.80 \pm 0.19$
&
$< 0.812 \times 10^{-23}$\\

Super-K I \cite{supk}
&
$\nu e$
&
$2.38\pm 0.02$
&
$<1.209 \times 10^{-23}$\\

Super-K II \cite{supk2}
&
$\nu e$
&
$2.41 \pm 0.05$
&
$< 1.189 \times 10^{-23}$\\

Super-K III \cite{supk3}
&
$\nu e$
&
$2.32 \pm 0.04$
&
$<1.321 \times 10^{-23}$\\

SNO Phase I \cite{sno1}
&
CC
&
$1.76^{+0.06}_{-0.05}$
&
$<1.536 \times 10^{-23}$\\

~~~~(pure $D_{2}O$)
&
$\nu e$
&
$2.39^{+0.24}_{-0.23}$
&
$<1.413 \times 10^{-23}$\\

&
NC
&
$5.09^{+0.44}_{-0.43}$
&
$<0.912 \times 10^{-23}$\\

SNO Phase II \cite{sno2}
&
CC
&
$1.68 \pm 0.06$
&
$<1.623 \times 10^{-23}$\\

~~~~(NaCl in $D_{2}O$)
&
$\nu e$
&
$2.35 \pm 0.22$
&
$<1.193 \times 10^{-23}$\\

&
NC
&
$4.94 \pm 0.21$
&
$< 0.798 \times 10^{-23}$\\

SNO Phase III \cite{aha08}
&
CC
&
$1.67^{+0.05}_{-0.04}$
&
$<1.415 \times 10^{-23}$\\

~~~~~($^{3}He$ counters)
&
$\nu e$
&
$1.77 ^{+0.24}_{-0.21}$
&
$<1.472 \times 10^{-23}$\\

&
NC
&
$5.54^{+0.33}_{-0.31}$
&
$<0.921 \times 10^{-23}$\\

Borexino \cite{borx}
&
$\nu e$
&
$2.4 \pm 0.4$
&
$<1.371 \times 10^{-23}$\\

\hline

SSM [$BPS08(GS)$] \cite{bps08}
&
-
&
$5.94(1\pm 0.11)$
&
-\\

SSM [$SHP11(GS)$] \cite{shp11}
&
-
&
$5.58(1\pm 0.14)$
&
-\\

\hline
\hline

\end{tabular}
\end{indented}
\end{center}
\end{table}

\section{Concluding Remarks}

In this paper, we have discussed and given analytical expression for phase of mass neutrino propagating along both null
and time-like geodesic in a general static spherically symmetric spacetime. Then we apply our phase
expression in three spherically symmetric solutions in different class of alternative gravity theories.
These phase expressions are being evaluated in the equatorial plane $\theta =\pi /2$ using spherical symmetry.
Using the phase expression we have calculated neutrino oscillation probability and hence put bounds on these
alternative theories. Thus this work not only shows an alternative way to constrain parameters in alternative
gravity theories but also shows neutrino phase and deviation from standard results through the quantities
$\delta l_{1}$ and $\delta l_{2}$ respectively. By setting different parameters to zero we have matched our
result to flat and Schwarzschild spacetime. We have also shown quiet generally that the phase along null
geodesic and time-like geodesic have a factor of $2$ in any general spherically symmetric spacetime.

In the last section we have presented the variation of oscillation length for introducing extra parameters
in our theory in comparison with flat and Schwarzschild spacetime. Also from the difference in oscillation
probability due to presence of alternative gravity can be used to constrain the parameters. For this
purpose we have used data on neutrino flux as measured by different experiments and then put bounds on
these parameters. The bounds on these parameters appear quiet small and hence not observed experimentally yet.
The best bound on the parameter $L$ appearing in the vacuum solution for $F(R)$ gravity is given from Table-1 as
$L^{-1}<1.895 \times 10^{-12}$. This is obtained using the SNO phase III results on solar neutrino oscillation.
Also the string tension parameter $\alpha ^{-1}$ in EMGB gravity has the best bounded value as $\alpha ^{-1}<5.759 \times 10^{-24}$
obtained from SNO phase II data on neutrino flux (see Table-2). In a similar manner we have the best bound on
$|\Lambda |$ as $|\Lambda| < 0.798 \times 10^{-23}$ for SNO phase II neutrino oscillation data (see Table-3).

However these results are derived for two generation neutrino oscillation, also in sun
only one type of neutrino is produced (namely, the electron type) and being of low energy we need not to consider all the
three generations. In recent times, there have been many works concerning ultra high energy neutrino
from AGN and other energetic astrophysical sources in the universe. So it would be quiet natural to discuss three generation neutrino
oscillation in gravitational field and apply the results of neutrino oscillation for these high energy neutrinos. Since these
neutrinos are generated in strong gravitational field, they may provide the behavior of gravity at such high field limit. We left these 
issues to be discuseed in some future work.

Also another interesting feature of this problem is the blueshift of neutrino phase. The oscillation length for
any spherically symmetric spacetime is proportional to local energy, interpreted as neutrino climbing
out of the gravitational potential well. However in this work we have used some solutions that are AdS
like on infinity thus decreasing the oscillation length. Thus our result is valid for any spherically
symmetric solution not only in four dimension but in any spacetime dimensions. Also the constraints
on the parameters are very interesting regarding their cosmological interpretation, which again makes
neutrino oscillation in curved spacetime a very interesting and profound problem in physics.
\ack
The author is funded by SPM fellowship from CSIR, Govt. of India. He thanks Chandan Hati, PRL, Ahmedabad and
Prof. Subenoy Chakraborty ,Jadavpur University for helpful discussions.
He is also thankful to the referees for helpful comments and suggestions that have helped to improve the manuscript.


\end{document}